\RequirePackage{amsmath} 
\documentclass[twocolumn,tighten]{aastex631}

\usepackage{mathtools}

\usepackage[capitalise,compress]{cleveref}
\newcommand{\crefrangeconjunction}{--}
\newcommand{\creflastconjunction}{, and\nobreakspace}
\crefname{equation}{Equation}{Equations}
\Crefname{equation}{Equation}{Equations}
\crefname{section}{Section}{Sections}
\Crefname{section}{Section}{Sections}
\crefname{figure}{Figure}{Figures}
\Crefname{figure}{Figure}{Figures}
\crefformat{subequations}{Equations~(#2#1#3)}
\Crefformat{subequations}{Equations~(#2#1#3)}

\makeatletter
\usepackage{etoolbox}
\patchcmd\H@refstepcounter{\protected@edef}{\protected@xdef}{}{}
\makeatother

\newcommand{\unit}[1]{\,\mathrm{#1}}
\newcommand{\diff}{\mathrm{d}}
\newcommand{\der}[2]{\frac{\diff#1}{\diff#2}}

\begin{document}


\title{Nonlinear coupling between magnetar QPOs}


\author{P. Pnigouras}
\affiliation{Departamento de F\'isica Aplicada, Universidad de Alicante, Campus de San Vicente del Raspeig, Alicante E-03690, Spain}

\author{S. K. Lander}
\affiliation{School of Engineering, Mathematics and Physics, University of East Anglia, Norwich NR4 7TJ, UK}


\begin{abstract}
    The quasi-periodic oscillations (QPOs) observed in the tails of magnetar giant $\gamma$-ray flares have long been interpreted as normal oscillation modes of these stars. However, most studies modelling QPOs have neglected some key features in the analyses of the signals, namely that QPOs appear to be detectable only intermittently and exhibit drifts in their frequencies. These are typical characteristics of nonlinear mode coupling, where, at leading order, the modes couple and evolve collectively as triplets. Using a representative triplet of modes, we solve the system's nonlinear equations of motion analytically and argue that the coupling is likely axial-axial-polar in nature, with the observed intermittence and frequency drifts providing a way to infer details of the magnetar's internal magnetic field geometry.
\end{abstract}




\section{Introduction} \label{sec:Introduction}

Two decades ago, \textit{quasi-periodic oscillations} (QPOs) were discovered \citep{israel05,StrohmayerWatts2005} in the aftermath of giant $\gamma$-ray flares from two different \emph{magnetars}, neutron stars (NSs) with very strong magnetic fields $B\sim 10^{14}-10^{15}\,\mathrm{G}$ characterised by their high-energy outburst activity. These QPOs, in the frequency range of tens to thousands of Hz, seem to represent the first detections of oscillation modes of isolated NSs and so should encode valuable information about their interior physics, including the critically important NS equation of state \citep{steiner_watts}, for which there is substantial uncertainty at high densities.

Although the lower-frequency QPOs had the expected properties of elastic shear modes of the NS crust, it quickly became clear that the problem was far more complex. The crust and core are threaded by a strong magnetic field that couples their dynamics, but whose broad structure is not well understood; this introduces an unwanted degree of ambiguity in the interpretation of observed QPOs as certain modes of the star. Another particularly concerning issue for QPO interpretation has been the possibility that the core admits a continuum of local oscillation frequencies rather than global oscillation modes, as this continuum would tend to drain energy from typical crustal modes \citep{levin07,vanhoven12}. Although certain additional pieces of physics appear to remove or alter the continuum---e.g. axisymmetric modes with a radial component \citep{sotani09,col-kok12}, non-axisymmetric modes \citep{LJP10,LJ11}, or the inclusion of neutron superfluidity \citep{gabler13,passland13,passland14}---these also further complicate attempts to identify particular observed QPOs with theoretically predicted modes.

Alongside modelling challenges, there have also been various re-analyses of the original data, resulting in conflicting conclusions about how significant and long-lived the different QPOs are \citep{hambaryan,pumpe18,miller19}. Although a few additional individual QPOs have been detected from various magnetars (e.g., \citealt{castro21}, \citealt{roberts23}), these are no substitute for one new set of high-quality data that reveals multiple QPOs, and modelling has reached an impasse of simply waiting passively for the next giant flare.

Almost all work on modelling QPOs has focussed on matching numbers: tuning the parameters of a magnetar model so that its oscillation-mode frequencies agree with  observed QPOs. Unfortunately, the theoretically predicted modes are sufficiently densely spaced in frequency that comparison with observations does not place robust constraints on the magnetar's underlying physics. Furthermore, this body of work ignores a major feature of the observations: the QPOs do not behave as if they were all excited at once and then simply decay, like the ringing of a bell. Instead, over hundreds of seconds following the giant flare, QPOs appear and disappear, suggesting some process of decay and re-excitation.

In this paper, we propose that these features are due to \textit{nonlinear couplings} between different modes. Modelling the couplings would not only help understand the full oscillation tails rather than just single modes, but could also lead to new insight about the physical origin of these modes.


\section{Revisiting the QPOs} \label{sec:Revisiting the QPOs}

To study magnetar mode coupling, we need to identify potential candidate QPOs that could be exhibiting this behaviour. We will look for QPOs following the giant flares of SGR 1806-20 and 1900+14 (hereafter SGR 1806 and SGR 1900) that were detected for more than just a very short segment, and preferably that are common to both magnetars and have been corroborated by more than one study. For SGR 1806, these criteria lead us to select the QPOs at 18, 26, 30, 93, and 150 Hz; there are also a few intriguing QPOs in the kHz range \citep{israel05,strohwatts06}. A subsequent search for QPOs from the far more frequent smaller bursts by \citet{huppen_short} revealed an additional QPO at 57 Hz, adding weight to the 59 Hz QPO found by \cite{hambaryan} in an earlier analysis of the giant-flare tail from the same magnetar. For SGR 1900, the original data is of lower quality, but the analysis of \citet{StrohmayerWatts2005} nonetheless yielded QPOs at 28, 53, 84, and 155 Hz.

Since mode coupling is known to cause \textit{frequency drift} (see  \cref{sec:Results}), we assume that QPOs of similar frequency represent the same underlying mode; i.e., for SGR 1806, we regard the strong 92.5 Hz and weak 95 Hz QPOs of \citet{israel05} and the 90 Hz QPO of \citet{strohwatts06} as the same mode; likewise for the 30 Hz \citep{israel05} and 26 Hz \citep{wattsstroh06} QPOs. This leaves us with four modes observed from both magnetars, with frequencies $\sim 28,\,55,\,88,\,150\unit{Hz}$. We will disregard the $\sim$150 Hz mode, partly because it is a very broad peak and partly because it is closer to the frequency range at which the core continuum of frequencies may affect results.
 
Our ansatz is that each of the above QPOs represents an oscillation mode of the star of \textit{magneto-elastic} character. To identify candidate physical modes, we use the fitting formulae from \citet[Equation 20]{GablerEtAl2016}. For the modes to have such a character and not be damped into the continuum, it is required that $10^{14}\,\mathrm{G}\lesssim B\lesssim 5\times 10^{15}\,\mathrm{G}$. We can also exclude any oscillations with non-constant phase, whose eigenfunction does not reach the stellar surface, from our modelling.

Satisfying the above, the only plausible triplet involves three of the ``$\,{}^l U_n$ modes$\,$'' of \citet{GablerEtAl2016}, namely
\begin{equation}
    {}^2 U_2\ ,\ {}^6 U_2\ ,\ {}^6 U_4. \label{Gabler best triplet}
\end{equation}
Varying the field strength so that the frequencies best match observations, we infer $B=8.6\times 10^{14}$ G for SGR 1806, which gives triplet mode frequencies of
\begin{equation}
    29.8,\ 56.2,\ 92.3\ \textrm{Hz}, \label{SGR 1806 triplet}
\end{equation}
and $B=7.2\times 10^{14}$ G for SGR 1900, yielding
\begin{equation}
    27.3,\ 53.7,\ 84.0\ \textrm{Hz}. \label{SGR 1900 triplet}
\end{equation}
These values of $B$ are in very good agreement with those inferred from spindown measurements: $7\times 10^{14}$ G for SGR 1900 \citep{mere06} and $8\times 10^{14}$ G for SGR 1806 \citep{kouveliotou98,younes17}. Note that the latter is often reported to have a substantially stronger field, $2\times 10^{15}$ G, but this value was inferred during a period of enhanced spindown around the time of the giant flare \citep{palmer05} -- likely due to an extended corona, making the dipole-spindown formula unreliable.


\section{Nonlinear mode-coupling theory} \label{sec:Nonlinear mode-coupling theory}

Nonlinear mode coupling is a result of the nonlinear interaction among the oscillation modes of a system. Ever since the work of \citet{Dziembowski1982}, this mechanism has been considerably studied in the context of astrophysics, mainly to explain the phenomenology of some variable stars \citep[e.g.,][]{WuGoldreich2001}, to study the saturation of NS instabilities \citep{SchenkEtAl2001,ArrasEtAl2003,PnigourasKokkotas2015,PnigourasKokkotas2016}, and to model the effects of nonlinear tides in binary systems \citep{WeinbergEtAl2012,WeinbergEtAl2013}.

The most basic system is that of three modes, which couple to each other at second order \citep[see, e.g.,][]{NayfehMook1979}. In the absence of dissipation, the equations of motion for this system may be written in terms of the mode amplitudes $Q_\alpha$:
\begin{subequations}
	\label[subequations]{amplitude equations of motion}
	\begin{align}
		\dot{Q}_\alpha & = i\omega_\alpha \, \kappa \, Q_\beta Q_\gamma e^{-i\Delta\omega t}, \label{amplitude equation of motion for mode alpha} \\
		\dot{Q}_\beta & = i\omega_\beta \, \kappa \, Q_\gamma^* Q_\alpha e^{i\Delta\omega t}, \label{amplitude equation of motion for mode beta} \\
		\dot{Q}_\gamma & = i\omega_\gamma \, \kappa \, Q_\alpha Q_\beta^* e^{i\Delta\omega t}, \label{amplitude equation of motion for mode gamma}
	\end{align}
\end{subequations}
where $\omega_\alpha$ is the mode frequency, $\Delta\omega=\omega_\alpha-\omega_\beta-\omega_\gamma$ is the \textit{detuning parameter} and $\kappa$ is the mode \textit{coupling coefficient}, defined as an overlap integral among the eigenfunctions $\xi_\alpha^i$ of the three modes (for an explicit expression of $\kappa$ see, e.g., \citealt{SchenkEtAl2001} or \citealt{Dziembowski1982}). The mode amplitudes $Q_\alpha$ are defined by the decomposition of a generic perturbation $\xi^i$ in terms of the modes as
\begin{equation}
    \xi^i(x^i,t)=\sum_\alpha Q_\alpha(t) \xi^i_\alpha(x^i) e^{i\omega_\alpha t} \label{mode decomposition}
\end{equation}
(note that $Q_\alpha$ does not contain the mode phase).

In spherical, i.e., non-rotating and unmagnetised, stars, each mode can be identified by a single spherical harmonic $Y_l^m(\theta,\phi)$ in a vector spherical harmonic decomposition, namely
\begin{equation}
    \xi^i=
        \frac{W}{r}Y_l^m\nabla^i r
        +V\nabla^i Y_l^m
        -iU\epsilon^{ijk}\nabla_j Y_l^m\nabla_k r.
    \label{mode spherical harmonic decomposition}
\end{equation}
Each mode is characterised by a multipole degree $l$, an azimuthal order $m$, and its overtone $n$, denoting the number of nodes in the mode's radial profile, described by the functions $W(r)$, $V(r)$, and $U(r)$.\footnote{The function $U$ here is not to be confused with the ${}^l U_n$ modes of the \citet{GablerEtAl2016} classification (see \cref{sec:Revisiting the QPOs}).} According to the standard mode classification, we have \textit{polar} modes, in which $U=0$, and \textit{axial} modes, in which $W=V=0$.

One may then show that
\begin{equation}
    \kappa \propto \iint Y_\alpha^* Y_\beta Y_\gamma \sin\theta\diff\theta\diff\phi, \label{coupling coefficient angular part}
\end{equation}
with $Y_\alpha\equiv Y_{l_\alpha}^{m_\alpha}$. This is a known integral \citep[e.g.,][]{SakuraiNapolitano2011}, which implies that $\kappa\ne 0$ if
\begin{gather}
    m_\alpha = m_\beta + m_\gamma \label{m selection rule} \\
    \shortintertext{and}
     l_a=l_b+l_c-2\lambda, \label{l selection rule}
\end{gather}
where $l_a\ge l_b\ge l_c$ and $\lambda=0,1,\ldots\lambda_\mathrm{max}\le l_c/2$, with the indices $a,b,c$ taking the values $\alpha,\beta,\gamma$, so that the mode $a$ has the largest degree and the mode $c$ has the lowest. \Cref{m selection rule,l selection rule} constitute the \textit{selection rules} that the coupled mode triplet has to satisfy and restrict the possible couplings. Furthermore, the equations of motion \eqref{amplitude equations of motion} imply that, among the many possible couplings, the triplets that ultimately affect the system's dynamics the most are those that satisfy the \textit{nonlinear resonance condition}
\begin{equation}
    \Delta\omega=\omega_\alpha-\omega_\beta-\omega_\gamma\simeq 0. \label{resonance condition}
\end{equation}

Note that the mode-coupling formalism laid out above is a result of second-order perturbation theory, meaning that these effects could, in principle, manifest themselves in any nonlinear numerical simulation. Apart from offering insight into the otherwise complex nonlinear dynamics, its main advantage is that it does not need to keep track of the mode phase, but rather of the slowly varying (due to nonlinear effects) mode amplitude $Q$, thus being able to achieve much longer integration times. This is necessary in any attempt to model the QPO features observed over long timescales.

Even though the potential relevance of nonlinear effects for QPOs has been suggested in the past \citep{strohwatts06}, previous efforts employing quadratic perturbations were not motivated by the same phenomena that interest us here. For instance, \citet{sotani24} derived the quadratically perturbed relativistic Euler equation of axisymmetric torsional (axial) modes in the NS crust and subsequently solved it numerically in the time domain, but their main result was to show the indirect excitation of normal modes, as well as the existence of additional modes, in the oscillation spectrum at second order, rather than attempting an explanation of the observed intermittence and frequency drifts of magnetar QPOs. Nonetheless, their calculation constitutes an important step toward a more realistic treatment of the problem and is worth building upon.


\section{Modelling the best triplet} \label{sec:Modelling the best triplet}

The magneto-elastic modes of \citet{GablerEtAl2016} depend on the magnetic field structure within the star, with nodes and maxima following the magnetic field lines. Thus, their indices $n$ and $l$ [see \cref{Gabler best triplet}] do not have the same meaning as above. An accurate description of such a mode would therefore require an infinite sum of $Y_l^m$ components in a vector spherical harmonic decomposition. This would present severe difficulties in a mode-coupling calculation, due to the various multipolar contributions from each mode entering the coupling coefficient \eqref{coupling coefficient angular part}. However, aiming for quantitatively accurate mode amplitudes in a calculation that is otherwise necessarily qualitative would be somewhat pointless. Instead, we argue that it is the similarity of the mode eigenfunctions involved in the nonlinear coupling that matters most, effectively establishing how good the overlap among them is, and hence, how strongly they are coupled. With this in mind, we will assume that the coupling of three magnetic ${}^l U_n$ modes is comparable to the coupling of three modes with eigenfunctions of the form of \cref{mode spherical harmonic decomposition} with the same $l,\, n$, and therefore proceed by representing each relevant mode from \citet{GablerEtAl2016} as 
\begin{equation}
    \xi^\phi_\alpha(r,\theta) = C\sin\left[\frac{(2n+1)\pi}{2}\frac{r}{R_*}\right]\frac{\partial_\theta Y_l^0}{\sin\theta}, \label{magnetic mode ansatz}
\end{equation}
where $R_*$ is the stellar radius and $C$ is a constant determined by the chosen normalisation. In the above we have adopted the form of an axisymmetric $(m=0)$ axial mode, in accordance with \mbox{\citet{GablerEtAl2016}}, and have also assumed that the mode's radial dependence is sinusoidal (see, e.g., \citealt{lee08}), with the requisite number of nodes $n$, and non-zero at the surface.

Like almost all other calculations of magnetar QPOs \citep[e.g.,][]{sam_and,sotani07,cerdaduran09}, the work of \citet{GablerEtAl2016} assumes axial oscillations, since these are more readily excited: they do not involve a radial displacement, which is inhibited by the composition-gradient stratification of the star \citep{duncan98}. We therefore begin with the result of \citet{SchenkEtAl2001}, who calculated the coupling coefficient for a triplet consisting of two axial modes $\beta,\,\gamma$, and a third mode $\alpha$ whose Eulerian density perturbation $\delta\rho$ vanishes:\\[1pt]
\begin{equation}
    \kappa_{\alpha\beta\gamma} = \frac{1}{2}\int p(\Gamma_1-\Gamma)\nabla_k\xi^{k*}_\alpha \, \nabla_i \xi^j_\beta \, \nabla_j \xi^i_\gamma\,\diff V, \label{coupling coefficient polar-axial-axial}
\end{equation}
with $\Gamma=\diff\ln p/\diff\ln\rho$ and $\Gamma_1=(\partial\ln p/\partial\ln\rho)_{x_\mathrm{p}}$, where $p$ is the pressure and $\rho$ is the density. The \textit{adiabatic exponent} $\Gamma_1$ differs from the background index $\Gamma$ if the chemical composition (here denoted by the proton fraction $x_\mathrm{p}$) changes throughout the star.

Since the divergence of an axial mode is zero, \cref{coupling coefficient polar-axial-axial} shows that there is no coupling between a triplet of axial modes. Thus, mode $\alpha$ must have a polar component. It may still have a leading-order axial piece, however; for instance, as is known from stellar perturbation theory, rotation introduces polar corrections to axial modes, and vice versa \citep[e.g.,][]{UnnoEtAl1989}. Therefore, only the polar piece of mode $\alpha$ contributes to $\nabla_i\xi^i_\alpha$. We will regard this divergence as a parameter to adjust; from \cref{mode spherical harmonic decomposition} we have $\nabla_i\xi^i_\alpha=f_\alpha Y_\alpha$, where
\begin{equation}
    f_\alpha=\frac{\zeta}{r^2}\left[\partial_r(rW_\alpha)-l_\alpha(l_\alpha+1) V_\alpha\right],
\end{equation}
with the small parameter $\zeta$ quantifying the relative amplitude of the polar piece to the axial piece of the mode.

There are various possible mechanisms for axial-polar mixing, but perhaps the most compelling is related to the magnetic field geometry. The purely poloidal background magnetic fields employed in \citet{GablerEtAl2016} and most other magnetar mode modelling are dynamically unstable \citep{markeytay73,wright73}, meaning that only a linked poloidal--toroidal geometry is realistic. Such a field geometry leads to axial modes mixing with polar modes in a manner proportional to the relative strength of toroidal to poloidal field \citep{col-kok12}. We will therefore regard $\zeta$ as a measure of the strength of this toroidal component.

Using \cref{mode spherical harmonic decomposition} and assuming axisymmetry, the axial-axial-polar coupling coefficient \eqref{coupling coefficient polar-axial-axial} takes the form
\begin{widetext}
    \begin{multline}
        \kappa_{\alpha\beta\gamma} = \frac{1}{4}
            \iint Y_\alpha^* Y_\beta Y_\gamma \sin\theta\diff\theta\diff\phi
            \int p(\Gamma_1-\Gamma)
                \left\{
                    f_\alpha\left(r g_\beta \partial_r g_\gamma+r \partial_r g_\beta g_\gamma\right) \left(-\Lambda_\alpha+\Lambda_\beta+\Lambda_\gamma\right)
                \vphantom{\left[\frac{(\Lambda_\alpha+\Lambda_\beta+\Lambda_\gamma)^2}{2}\right]}\right. \\
                \left.
                    + f_\alpha g_\beta g_\gamma \left[\frac{(\Lambda_\alpha+\Lambda_\beta+\Lambda_\gamma)^2}{2}-(\Lambda_\alpha^2+\Lambda_\beta^2+\Lambda_\gamma^2)+3(-\Lambda_\alpha+\Lambda_\beta+\Lambda_\gamma)\right]
                \right\} r^2 \diff r, \label{coupling coefficient polar-axial-axial explicit}
    \end{multline}
\end{widetext}
where $\Lambda=l(l+1)$ and $g = U/r^2$. For the evaluation of the functions $W$, $V$, and $U$, we will assume that the polar components $\xi^r$ and $\xi^\theta$ of mode $\alpha$ take on the same radial dependence as the axial component $\xi^\phi$, according to the ansatz \eqref{magnetic mode ansatz}.

We also define the mode energy as $E_\alpha=2 I_\alpha\omega_\alpha^2|Q_\alpha|^2$, where
\begin{equation}
    I=\int\xi^{i*}\xi_i\rho\diff V
\end{equation}
\citep[see, e.g.,][]{SchenkEtAl2001}. The modes will be normalised by fixing their energy to unit amplitude, i.e., $2 I_\alpha\omega_\alpha^2=Mc^2$, with $M$ being the mass of the star and $c$ being the speed of light. Hence, the mode amplitude is simply related to the mode energy through $E_\alpha=|Q_\alpha|^2 Mc^2$.

The same normalisation has been tacitly assumed for the coupling coefficient $\kappa$ in the derivation of \cref{amplitude equations of motion}; $\kappa$ is calculated in units of the chosen energy value---in this case $Mc^2$---so the expression of \cref{coupling coefficient polar-axial-axial explicit} needs to be divided by that before being used in \cref{amplitude equations of motion}. An additional subtlety is that the modes $\beta$ and $\gamma$ also ought to have polar components of order $\zeta$, which would thus couple to the axial components of the other two modes in the triplet (we will, however, neglect couplings of order $\zeta^2$ and higher, involving the polar components of two or three modes). This means that we also need to evaluate $\kappa_{\bar{\beta}\gamma\bar{\alpha}}$ and $\kappa_{\bar{\gamma}\bar{\alpha}\beta}$, with the bar over the mode label meaning that the mode eigenfunction has to be complex conjugated in \cref{coupling coefficient polar-axial-axial explicit}. The complete expression for the coupling coefficient is then
\begin{equation}
    \kappa=\frac{1}{Mc^2}\left(\kappa_{\alpha\beta\gamma}+\kappa_{\bar{\beta}\gamma\bar{\alpha}}+\kappa_{\bar{\gamma}\bar{\alpha}\beta}\right).
\end{equation}

We will assume a background star with $M=1.4\,M_\odot$ ($M_\odot$ being the solar mass) and $R_*=10\unit{km}$, described by a polytropic equation of state $p\propto\rho^\Gamma$, with $\Gamma=2$, for which the background quantities may be obtained analytically, and set the adiabatic exponent $\Gamma_1=2.1$ \citep[see, for example,][]{AnderssonPnigouras2020}.


\section{Results} \label{sec:Results}

The coupled equations of motion \eqref{amplitude equations of motion} admit an analytical solution. Decomposing the complex mode amplitude $Q_\alpha$ into its real amplitude and phase components as
\begin{equation}
    Q_\alpha=\frac{\varepsilon_\alpha e^{i\vartheta_\alpha}}{\kappa\sqrt{\omega_\beta\omega_\gamma}}\, ,\label{real amplitude and phase variables}
\end{equation}
\cref{amplitude equations of motion} become
\begin{subequations}
	\label[subequations]{amplitude equations of motion with amplitude and phase variables}
	\begin{gather}
		\dot{\varepsilon}_\alpha = \varepsilon_\beta\varepsilon_\gamma\sin\varphi, \label{amplitude equation of motion with amplitude and phase variables for mode alpha} \\
		\dot{\varepsilon}_\beta = -\varepsilon_\gamma\varepsilon_\alpha\sin\varphi, \label{amplitude equation of motion with amplitude and phase variables for mode beta} \\
		\dot{\varepsilon}_\gamma = -\varepsilon_\alpha\varepsilon_\beta\sin\varphi, \label{amplitude equation of motion with amplitude and phase variables for mode gamma} \\
        \shortintertext{and}
        \dot{\varphi} = \cos\varphi\left(\frac{\varepsilon_\beta\varepsilon_\gamma}{\varepsilon_\alpha}-\frac{\varepsilon_\gamma\varepsilon_\alpha}{\varepsilon_\beta}-\frac{\varepsilon_\alpha\varepsilon_\beta}{\varepsilon_\gamma}\right)+\Delta\omega, \label{phase equation of motion}
	\end{gather}
\end{subequations}
where $\varphi=\vartheta_\alpha-\vartheta_\beta-\vartheta_\gamma+\Delta\omega t$.

Including the phase $\vartheta_\alpha$ in the harmonic time dependence of the mode, we get $\xi^i_\alpha\propto e^{i(\omega_\alpha t+\vartheta_\alpha)}$, suggesting that the nonlinear coupling induces modulation of the mode frequency of the form $\omega_\alpha'=\omega_\alpha+\dot{\vartheta}_\alpha$, with the \textit{shift} $\dot{\vartheta}_\alpha$ given by
\begin{equation}
    \dot{\vartheta}_\alpha=\frac{\varepsilon_\beta \varepsilon_\gamma}{\varepsilon_\alpha}\cos\varphi. \label{eigenfrequency shift}
\end{equation}

Upon inspection, from the equations of motion \eqref{amplitude equations of motion with amplitude and phase variables} we can obtain the following constants of integration:
\begin{gather}
    \mathcal{E}_1 = \varepsilon_\alpha^2+\varepsilon_\beta^2, \label{integration constant Epsilon1} \\
    \mathcal{E}_2 = \varepsilon_\alpha^2+\varepsilon_\gamma^2, \label{integration constant Epsilon2} \\
    \shortintertext{and}
    L = \varepsilon_\alpha\varepsilon_\beta\varepsilon_\gamma\cos\varphi+\frac{\Delta\omega}{2}\varepsilon_\alpha^2\, , \label{integration constant L1}
\end{gather}
with $\mathcal{E}_1$ and $\mathcal{E}_2$ being related to the energy of the mode triplet, whereas $L$ dictates the evolution of the frequency shifts $\dot{\vartheta}_\alpha$. Combining \cref{amplitude equation of motion with amplitude and phase variables for mode alpha,integration constant L1}, and making the replacement $\varepsilon_\alpha^2=q\mathcal{E}_1$, we get
\begin{equation}
    \left(\der{q}{t}\right)^2=4\mathcal{E}_1\left[q (1-q)(\mu-q)-\frac{1}{\mathcal{E}_1}\left(\frac{\Delta\omega}{2}q-\frac{L}{\mathcal{E}_1}\right)^2\right], \label{amplitude equation of motion in terms of q for mode alpha}
\end{equation}
with $\mu=\mathcal{E}_2/\mathcal{E}_1$. The expression in brackets on the right-hand side of \cref{amplitude equation of motion in terms of q for mode alpha} is a cubic polynomial with roots $q_1,\,q_2$ and $q_3$, such that $q\in[q_1,\,q_2\leqslant 1]$ and $q_3\geqslant 1$.

We further replace $q=q_1+(q_2-q_1)\sin^2\delta$, which makes \cref{amplitude equation of motion in terms of q for mode alpha}
\begin{equation}
     \left(\der{\delta}{t}\right)^2=\mathcal{E}_1(q_3-q_1)\left(1-k^2\sin^2\delta\right), \label{amplitude equation of motion in terms of delta factorised for mode alpha}
\end{equation}
with $k^2=(q_2-q_1)/(q_3-q_1)\leqslant 1$. We integrate to get
\begin{equation}
    \int_{\delta_0}^\delta\frac{\diff\delta'}{\sqrt{1-k^2\sin^2\delta'}}=\sqrt{\mathcal{E}_1(q_3-q_1)}t, \label{amplitude equation of motion in terms of delta factorised integrated for mode alpha}
\end{equation}
from which we can define the \textit{Jacobi elliptic function} $\mathrm{sn}\left(u,k\right)=\sin\delta$ \citep[e.g.,][]{AbramowitzStegun1972}, where
\begin{equation}
    u(t)=\sqrt{\mathcal{E}_1(q_3-q_1)}t+\int_0^{\delta_0}\frac{\diff\delta}{\sqrt{1-k^2\sin^2\delta}}\, , \label{Jacobi elliptic function argument u}
\end{equation}
with the angle $\delta_0$ determined by the initial conditions for the mode amplitudes.

Switching to our initial variables, the solution to \cref{amplitude equations of motion} is
\begin{subequations}
	\label[subequations]{amplitude equations of motion solutions}
	\begin{gather}
		|Q_\alpha|^2 = \frac{\mathcal{E}_1}{\omega_\beta\omega_\gamma\kappa^2}\left[q_1+(q_2-q_1)\mathrm{sn}^2(u,k)\right], \label{amplitude equation of motion with amplitude and phase variables solution for mode alpha} \\
		|Q_\beta|^2 = \frac{\mathcal{E}_1}{\omega_\gamma\omega_\alpha\kappa^2}\left[1-q_1-(q_2-q_1)\mathrm{sn}^2(u,k)\right], \label{amplitude equation of motion with amplitude and phase variables solution for mode beta} \\
		|Q_\gamma|^2 = \frac{\mathcal{E}_1}{\omega_\alpha\omega_\beta\kappa^2}\left[\mu-q_1-(q_2-q_1)\mathrm{sn}^2(u,k)\right]. \label{amplitude equation of motion with amplitude and phase variables solution for mode gamma}
	\end{gather}
\end{subequations}
These are periodic with respect to time, with the period given by
\begin{equation}
    T=\frac{2 K(k)}{\sqrt{\mathcal{E}_1(q_3-q_1)}}, \label{amplitude modulation period}
\end{equation}
$K(k)$ being the \textit{complete elliptic integral of the first kind}.

To demonstrate the system's behaviour, we will use the mode triplet of SGR 1900 [\cref{SGR 1900 triplet}], with $\alpha$ labelling the high-frequency \textit{(parent)} mode and $\beta,\,\gamma$ the mid- and low-frequency \textit{(daughter)} modes, respectively. The results are similar for the mode triplet in SGR 1806 [\cref{SGR 1806 triplet}]. The initial mode amplitudes $|Q|$ are fixed, leaving $\zeta$ as the only free parameter.

\begin{figure}
    \centering
    \includegraphics[width=\columnwidth]{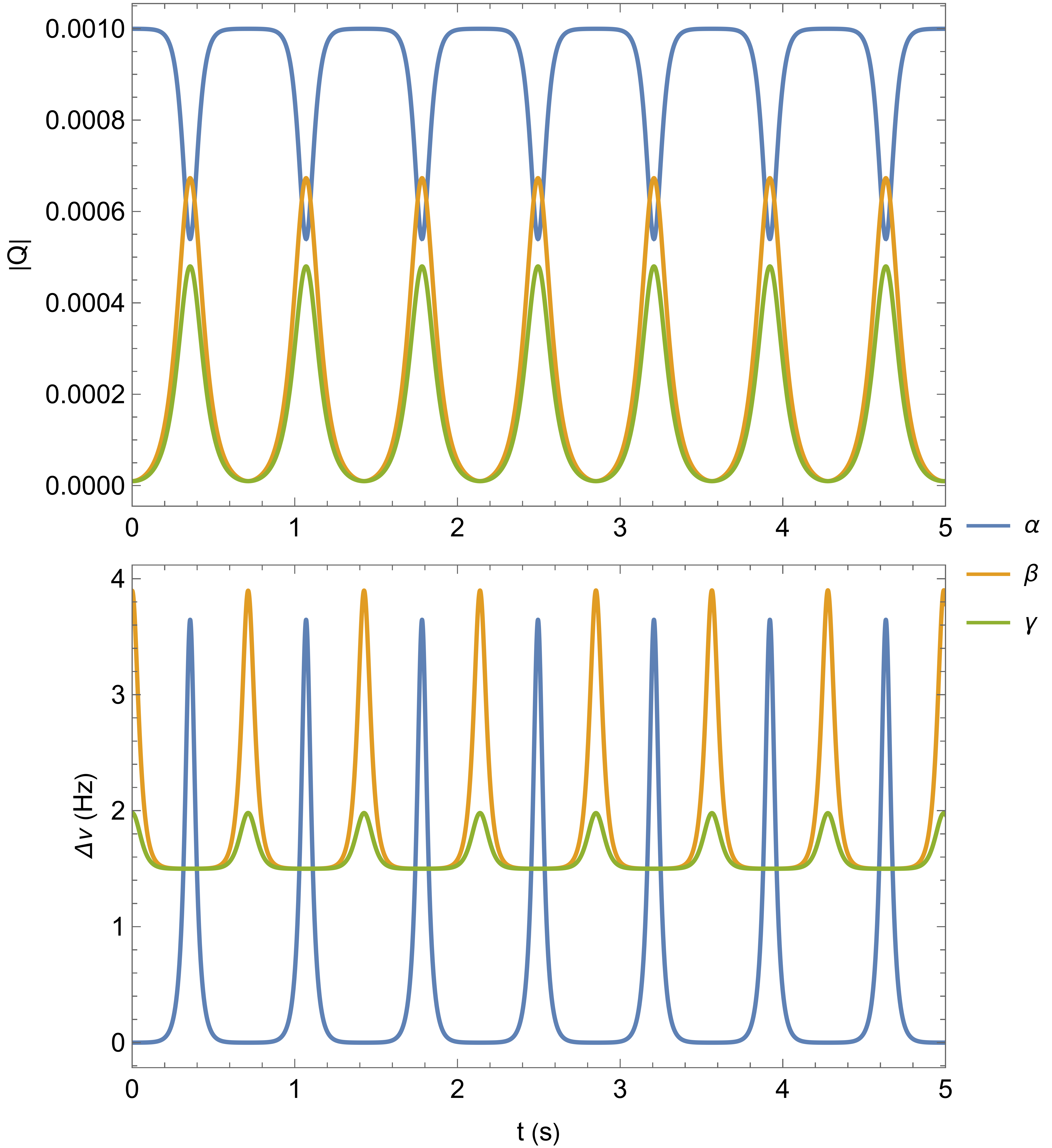}
    \caption{Evolution of our representative coupled mode triplet, with $\alpha$ labelling the high-frequency (parent) mode and $\beta,\,\gamma$ the mid- and low-frequency (daughter) modes, respectively. We plot the mode amplitude $|Q|$ \textit{(top)} and mode frequency shift $\varDelta\nu\equiv\dot{\vartheta}/2\pi$ in Hz \textit{(bottom)} as functions of time, for $\zeta=4\times 10^{-4}$ and initial conditions $|Q_\alpha|=10^{-3},|Q_\beta|=|Q_\gamma|=10^{-5}$.}
    \label{fig:triplet demonstration evolution}
\end{figure}

\begin{figure*}
    \centering
    \includegraphics[width=\textwidth]{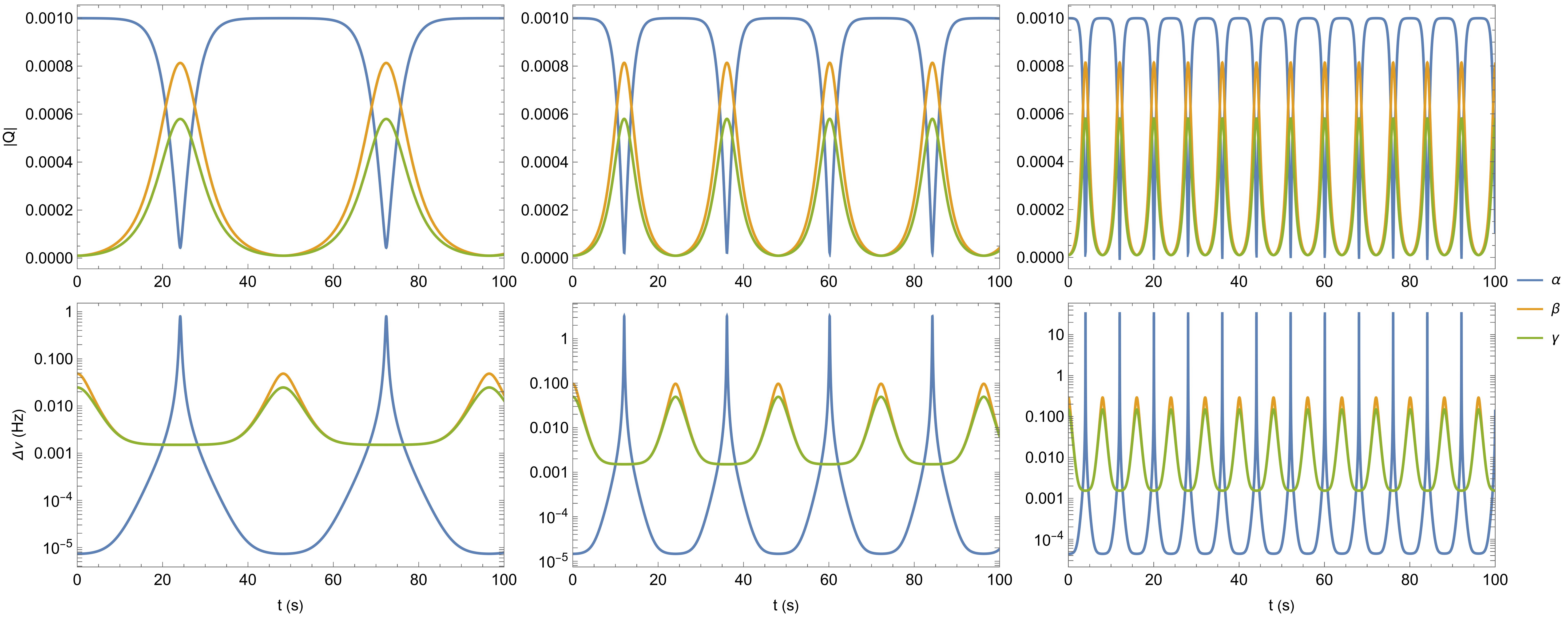}
    \caption{The effect of varying $\zeta$ on $|Q|$ \textit{(top panels)} and $\varDelta\nu$ \textit{(bottom panels)}, for the same initial conditions as for \cref{fig:triplet demonstration evolution}, but with the detuning dropped by a factor of 1000. From left to right: $\zeta=5\times 10^{-6},\, 10^{-5},\, 3\times 10^{-5}$.}
    \label{fig:triplet amplitude and frequency shift evolution}
\end{figure*}

In \cref{fig:triplet demonstration evolution} we plot the evolution of the mode amplitudes $|Q|$ and frequency shifts $\varDelta\nu\equiv\dot{\vartheta}/2\pi$ (in Hz) for $\zeta=4\times 10^{-4}$ and initial conditions $|Q_\alpha|=10^{-3}$, $|Q_\beta|=|Q_\gamma|=10^{-5}$. The energy exchange between the modes, characteristic for these systems, occurs periodically $(T\approx 0.7\unit{s})$, with the high-energy mode pumping energy into the low-energy modes, which then transfer it back, and the cycle repeats. Likewise, the frequency shift also oscillates periodically, with its maxima corresponding to minima in the amplitude and reaching values up to 4 Hz.

The period $T$ of the amplitude modulation and the magnitude of the frequency shifts can be adjusted by changing $\zeta$. Unless $\zeta$ falls below a certain limit value $\zeta_\mathrm{lim}$ (see below), $(q_3-q_1)\sim 1$ and $K(k)\sim 1$, so from \cref{amplitude modulation period} we have $T\sim\mathcal{E}_1^{-1/2}\propto\kappa^{-1}\propto\zeta^{-1}$. The scaling of $\varDelta\nu$ with $\zeta$ is more complicated. Expressing \cref{integration constant L1} as $L=\varepsilon_\alpha^2(\dot{\vartheta}_\alpha+\Delta\omega/2)$ and noticing that $\varepsilon_\alpha^2\dot{\vartheta}_\alpha=\varepsilon_\beta^2\dot{\vartheta}_\beta=\varepsilon_\gamma^2\dot{\vartheta}_\gamma$, one may derive the magnitudes of the frequency shifts as
\begin{subequations}
    \label{eigenfrequency shifts magnitudes}
    \begin{gather}
        \dot{\vartheta}_\alpha^\mathrm{mag}=\frac{L}{\mathcal{E}_1}\frac{q_2-q_1}{q_1 q_2}, \label{eigenfrequency shift magnitude for mode alpha} \\
        \dot{\vartheta}_\beta^\mathrm{mag}=\left|\frac{\Delta\omega}{2}-\frac{L}{\mathcal{E}_1}\right|\frac{q_2-q_1}{(1-q_1)(1-q_2)}, \label{eigenfrequency shift magnitude for mode beta} \\
        \dot{\vartheta}_\gamma^\mathrm{mag}=\left|\mu\frac{\Delta\omega}{2}-\frac{L}{\mathcal{E}_1}\right|\frac{q_2-q_1}{(\mu-q_1)(\mu-q_2)}.\label{eigenfrequency shift magnitude for mode gamma}
    \end{gather}
\end{subequations}
These are increasing functions of $\zeta$ and approach zero in the linear limit, when $\zeta\approx\zeta_\mathrm{lim}$ (or $q_1\approx q_2$; see below). Hence, decreasing $\zeta$ leads both to larger modulation periods and smaller frequency shifts.

There is, however, a limit value, $\zeta_\mathrm{lim}$, to which we alluded earlier, below which the modes are too weakly coupled and their amplitudes are practically constant (essentially, the system approaches the linear limit). This occurs when $q_1\approx q_2$, or when the discriminant of the cubic polynomial in $q$ from \cref{amplitude equation of motion in terms of q for mode alpha} gets close to zero. The value of $\zeta_\mathrm{lim}$ depends on the value of the detuning $\Delta\omega$: the larger $\Delta\omega$ is, the larger $\zeta_\mathrm{lim}$ is. This is sensible, because a small detuning implies a stronger nonlinear resonance and, hence, stronger nonlinear coupling among the three modes [see \cref{resonance condition}], which allows for smaller values of the coupling coefficient.

For the triplet we are examining, the case depicted in \cref{fig:triplet demonstration evolution} is already near $\zeta_\mathrm{lim}$. In order to explain the appearance and disappearance of modes over timescales of order $10-100\unit{s}$, which is what QPO observations suggest, we need lower values of $\zeta$. Let us, therefore, reduce the original value of the detuning $(\Delta\omega/2\pi=3\unit{Hz})$ by a factor of 1000. For this case, we plot $|Q|$ and $\varDelta\nu$ in \cref{fig:triplet amplitude and frequency shift evolution} for three different values of $\zeta$ and using the same initial conditions as before, in order to demonstrate that the variation of $\zeta$ affects $T$ and $\varDelta\nu$ according to the scalings presented before. In particular, for $\zeta=5\times 10^{-6}$ we obtain $T\approx 50\unit{s}$, and $\varDelta\nu_\alpha$ reaches values up to 1 Hz; for $\zeta=10^{-5}$ the period drops to $T\approx 25\unit{s}$, and $\varDelta\nu_\alpha$ increases to up to 3 Hz; and for $\zeta=3\times 10^{-5}$ the period is $T\approx 9\unit{s}$, with $\varDelta\nu_\alpha$ getting up to 30 Hz.

\begin{figure}
    \centering
    \includegraphics[width=\columnwidth]{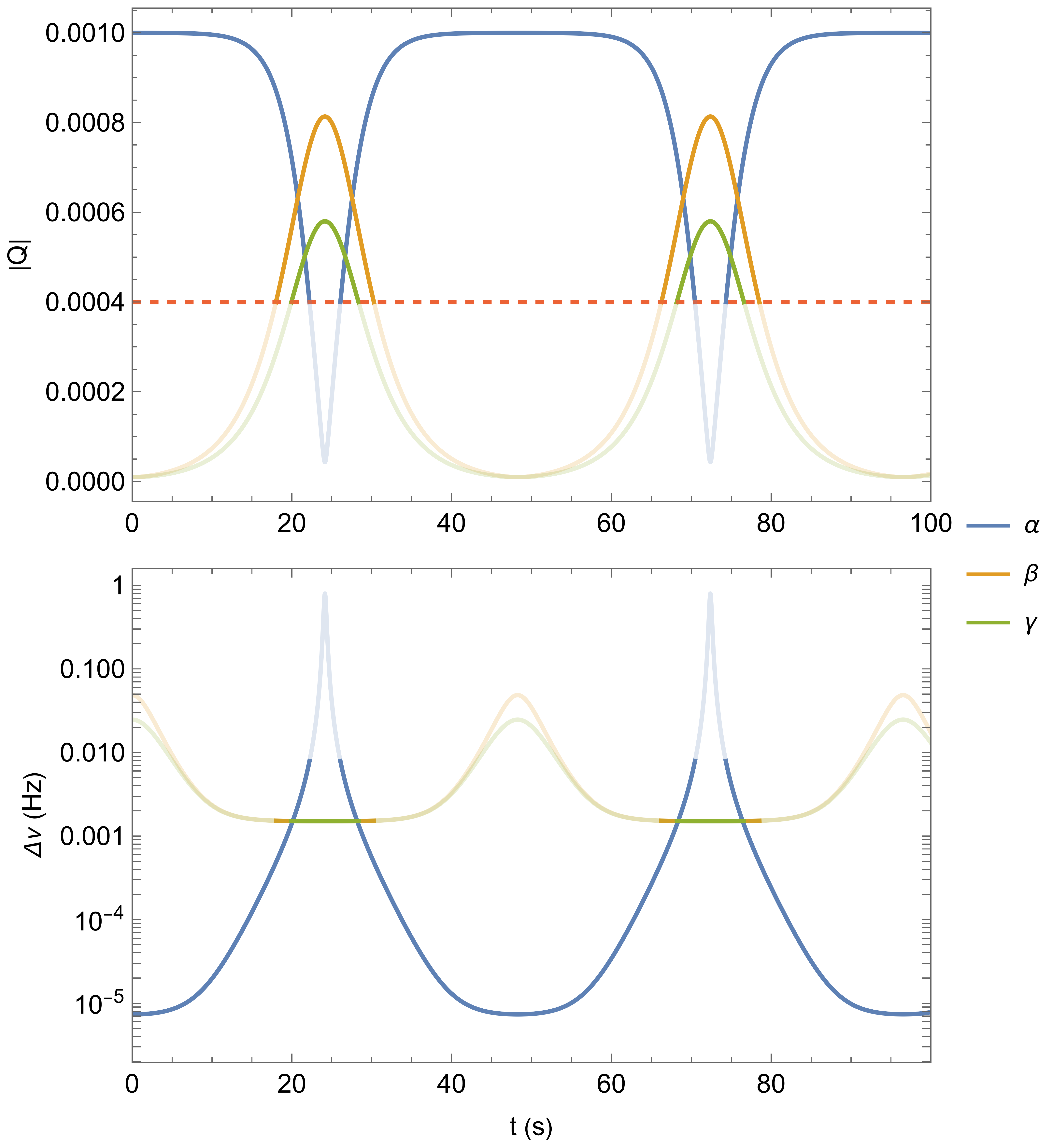}
    \caption{Evolution of $|Q|$ and $\varDelta\nu$ from the first column of \cref{fig:triplet amplitude and frequency shift evolution}, but now assuming an amplitude detection cutoff $|Q|=4\times 10^{-4}$ \textit{(dashed line)}.}
    \label{fig:triplet amplitude and frequency shift evolution with amplitude cutoff}
\end{figure}

Moreover, in \cref{fig:triplet amplitude and frequency shift evolution with amplitude cutoff} we plot $|Q|$ and $\varDelta\nu$ for the same parameters as in the first column of \cref{fig:triplet amplitude and frequency shift evolution}, but also assuming an amplitude detection cutoff, set to $|Q|=4\times 10^{-4}$. When the amplitude becomes lower than this cutoff value, the mode is regarded as undetectable, and its amplitude and frequency shift curves are plotted as semi-transparent. This way we mimic the observed appearance and disappearance of modes in giant-flare tails.

Finally, note that our choice of initial conditions is such that $\zeta_\mathrm{lim}\ll 1$. Increasing the initial mode amplitudes practically has the same effect as increasing $\zeta$ (and vice versa), which is to be expected; a large mode amplitude would imply that we are deep into the nonlinear regime, hence the coupling coefficient (parametrised by $\zeta$) need not be as large for the characteristic energy exchange to occur among the three modes.


\section{Discussion} \label{sec:Discussion}

The discovery of magnetar QPOs triggered a concerted effort to develop theoretical models of oscillating NSs, with the intention of identifying the observed QPOs as specific oscillation modes of the model magnetar. This work was, ultimately, motivated by the prospect of using QPO identification as a way to probe unknown aspects of the interior physics of magnetars: their magnetic field strength, crustal properties, and the equation of state of matter beyond nuclear density \citep{steiner_watts,sotani12,gabler_explore_B}. Although this endeavour has not been as fruitful as first hoped, due both to the limited quality of the data and to the complexity of the theoretical modelling problem, the advances made to date will be invaluable in the interpretation of QPOs following any future giant flare. Furthermore, this work has since found a separate application, related to QPOs associated with $\gamma$-ray bursts (GRBs). One class of GRBs is believed to be related to compact-object binary mergers, and QPOs were recently detected in a GRB precursor event \citep{xiao24,chir24}, providing insights into the pre-merger physics \citep{suv22}.

The two sets of magnetar-QPO data represent more than just a set of numbers to be matched, however. In this paper we revisit the original X-ray tails following the giant flares of SGR 1806 and SGR 1900, arguing that these hold additional valuable information that deserves study. In particular, the apparent disappearance and reappearance of magnetar QPOs, together with their drifting frequencies, are important observations that have previously received little attention from theoretical modelling (a rare exception being the work of \citealt{levin07}). We have found that both phenomena can be explained through nonlinear mode coupling. In such a system, mode amplitudes have oscillatory variation in time (rather than simply decaying), and also time-varying frequency shifts, providing a concrete physical reason to believe that, e.g., the 26 and 30 Hz QPOs of SGR 1806 are the same oscillation mode. Despite earlier recognition of the potential importance of nonlinear effects in QPO modelling, e.g. by \citet{sotani24}, where the second-order axial oscillation spectrum of the NS crust was derived, their capability of explaining these QPO features had not been appreciated before.

There is one key free parameter $\zeta$ in our coupled mode triplet, which quantifies the strength of the subdominant polar piece of a predominantly axial mode. This is required, because the most readily excited magnetar modes are axial, but three purely axial modes do not experience any coupling. The value of $\zeta$ dictates the timescale on which a given mode dis-/re-appears, and also its drift in frequency. We argue that it is natural to associate $\zeta$ with the toroidal component of the star's magnetic field; the behaviour of a magnetar's QPOs could thus be one of very few possible mechanisms for inferring the star's internal field geometry.

Although we can mimic QPO dis-/re-appearance over the observed timescales, or observed shifts in frequency of some QPOs, it is more challenging to get both right with our simple coupled-triplet model---partly because we predict the maximum frequency shift to occur when the mode is least likely to be detectable. Furthermore, the real data does not indicate that QPO variability is strictly periodic, as our model predicts. It is, however, easy to see where further complexity could come from in the mode coupling, perhaps resulting in better agreement with observations. Firstly, the fact that the $\sim 57,\sim 90,\sim 150$ Hz QPOs are also likely to couple together (based on the frequency selection rule, i.e. $57+90\approx 150$)---involving two of the same modes as our fiducial triplet here---suggests a possible coupling between the two \emph{triplets}. Further complexity would come if we allowed for energy to be dissipated from the modes (rather than just transferred between them). Finally, we need to recall that we do not observe these QPOs directly, as oscillations of the star's surface, but rather as modulations to the brightness of a fireball anchored to the star by magnetic field lines \citep{gabler14,akgun18}. This process will affect how `visible' certain modes are, on top of the intrinsic changes in amplitude that mode coupling gives.

Higher-frequency QPOs $(\gtrsim 600\unit{Hz})$ are more likely to represent non-axisymmetric modes, whose eigenfunctions would be sufficiently different as to inhibit coupling with axisymmetric mode triplets like those we focus on here. In the tail following  the giant flare of SGR 1806, these high-frequency QPOs exhibited similar behaviour to the lower-frequency QPOs \citep{wattsstroh06}, suggesting a process of mode coupling in this case too. Intriguingly, high-frequency (kHz) QPOs were also seen \emph{during} the giant flare from an extragalactic magnetar \citep{castro21}, providing an additional probe of the underlying physics driving this phenomenon \citep{sotani_solo_24} and suggesting that these kHz oscillations may be preferentially excited during the initial phase, perhaps coupling to other modes much later on.


\begin{acknowledgments}
    P.P. acknowledges support from the Mar\'ia Zambrano Fellowship Programme (ZAMBRANO21-22), funded by the Spanish Ministry of Universities and the University of Alicante through the European Union's ``Next Generation EU'' package, as well as from the grant PID2021-127495NB-I00, funded by MCIN/AEI/10.13039/501100011033 and by the European Union, from the Astrophysics and High Energy Physics programme of the Generalitat Valenciana ASFAE/2022/026, funded by the Spanish Ministry of Science and Innovation (MCIN) and the European Union's ``Next Generation EU'' package (PRTR-C17.I1), and from the Prometeo 2023 excellence programme grant CIPROM/2022/13, funded by the Ministry of Education, Culture, Universities, and Occupation of the Generalitat Valenciana.
\end{acknowledgments}




\bibliography{references}

\begin{thebibliography}{}
\expandafter\ifx\csname natexlab\endcsname\relax\def\natexlab#1{#1}\fi
\providecommand{\url}[1]{\href{#1}{#1}}
\providecommand{\dodoi}[1]{doi:~\href{http://doi.org/#1}{\nolinkurl{#1}}}
\providecommand{\doeprint}[1]{\href{http://ascl.net/#1}{\nolinkurl{http://ascl.net/#1}}}
\providecommand{\doarXiv}[1]{\href{https://arxiv.org/abs/#1}{\nolinkurl{https://arxiv.org/abs/#1}}}

\bibitem[{{Abramowitz} \& {Stegun}(1972)}]{AbramowitzStegun1972}
{Abramowitz}, M., \& {Stegun}, I.~A. 1972, {H}andbook of {M}athematical
  {F}unctions (New York: Dover).
\newblock \url{http://adsabs.harvard.edu/abs/1972hmfw.book.....A}

\bibitem[{{Akg{\"u}n} {et~al.}(2018){Akg{\"u}n}, {Cerd{\'a}-Dur{\'a}n},
  {Miralles}, \& {Pons}}]{akgun18}
{Akg{\"u}n}, T., {Cerd{\'a}-Dur{\'a}n}, P., {Miralles}, J.~A., \& {Pons}, J.~A.
  2018, \mnras, 481, 5331, \dodoi{10.1093/mnras/sty2669}

\bibitem[{{Andersson} \& {Pnigouras}(2020)}]{AnderssonPnigouras2020}
{Andersson}, N., \& {Pnigouras}, P. 2020, \prd, 101, 083001,
  \dodoi{10.1103/PhysRevD.101.083001}

\bibitem[{{Arras} {et~al.}(2003){Arras}, {Flanagan}, {Morsink}, {Schenk},
  {Teukolsky}, \& {Wasserman}}]{ArrasEtAl2003}
{Arras}, P., {Flanagan}, {\'E}.~{\'E}., {Morsink}, S.~M., {et~al.} 2003, \apj,
  591, 1129, \dodoi{10.1086/374657}

\bibitem[{{Castro-Tirado} {et~al.}(2021){Castro-Tirado}, {{\O}stgaard},
  {G{\"o}{\u{g}}{\"u}{\c{s}}}, {S{\'a}nchez-Gil}, {Pascual-Granado}, {Reglero},
  {Mezentsev}, {Gabler}, {Marisaldi}, {Neubert}, {Budtz-J{\o}rgensen},
  {Lindanger}, {Sarria}, {Kuvvetli}, {Cerd{\'a}-Dur{\'a}n},
  {Navarro-Gonz{\'a}lez}, {Font}, {Zhang}, {Lund}, {Oxborrow}, {Brandt},
  {Caballero-Garc{\'\i}a}, {Carrasco-Garc{\'\i}a}, {Castell{\'o}n}, {Castro
  Tirado}, {Christiansen}, {Eyles}, {Fern{\'a}ndez-Garc{\'\i}a}, {Genov},
  {Guziy}, {Hu}, {Nicuesa Guelbenzu}, {Pandey}, {Peng}, {P{\'e}rez del Pulgar},
  {Reina Terol}, {Rodr{\'\i}guez}, {S{\'a}nchez-Ram{\'\i}rez}, {Sun},
  {Ullaland}, \& {Yang}}]{castro21}
{Castro-Tirado}, A.~J., {{\O}stgaard}, N., {G{\"o}{\u{g}}{\"u}{\c{s}}}, E.,
  {et~al.} 2021, \nat, 600, 621, \dodoi{10.1038/s41586-021-04101-1}

\bibitem[{{Cerd{\'a}-Dur{\'a}n} {et~al.}(2009){Cerd{\'a}-Dur{\'a}n},
  {Stergioulas}, \& {Font}}]{cerdaduran09}
{Cerd{\'a}-Dur{\'a}n}, P., {Stergioulas}, N., \& {Font}, J.~A. 2009, \mnras,
  397, 1607, \dodoi{10.1111/j.1365-2966.2009.15056.x}

\bibitem[{{Chirenti} {et~al.}(2024){Chirenti}, {Dichiara}, {Lien}, \&
  {Miller}}]{chir24}
{Chirenti}, C., {Dichiara}, S., {Lien}, A., \& {Miller}, M.~C. 2024, \apj, 967,
  26, \dodoi{10.3847/1538-4357/ad3bb7}

\bibitem[{{Colaiuda} \& {Kokkotas}(2012)}]{col-kok12}
{Colaiuda}, A., \& {Kokkotas}, K.~D. 2012, \mnras, 423, 811,
  \dodoi{10.1111/j.1365-2966.2012.20919.x}

\bibitem[{{Duncan}(1998)}]{duncan98}
{Duncan}, R.~C. 1998, \apjl, 498, L45, \dodoi{10.1086/311303}

\bibitem[{{Dziembowski}(1982)}]{Dziembowski1982}
{Dziembowski}, W. 1982, Acta Astron., 32, 147.
\newblock \url{http://adsabs.harvard.edu/abs/1982AcA....32..147D}

\bibitem[{{Gabler} {et~al.}(2013{\natexlab{a}}){Gabler}, {Cerd{\'a}-Dur{\'a}n},
  {Font}, {M{\"u}ller}, \& {Stergioulas}}]{gabler_explore_B}
{Gabler}, M., {Cerd{\'a}-Dur{\'a}n}, P., {Font}, J.~A., {M{\"u}ller}, E., \&
  {Stergioulas}, N. 2013{\natexlab{a}}, \mnras, 430, 1811,
  \dodoi{10.1093/mnras/sts721}

\bibitem[{{Gabler} {et~al.}(2013{\natexlab{b}}){Gabler}, {Cerd{\'a}-Dur{\'a}n},
  {Stergioulas}, {Font}, \& {M{\"u}ller}}]{gabler13}
{Gabler}, M., {Cerd{\'a}-Dur{\'a}n}, P., {Stergioulas}, N., {Font}, J.~A., \&
  {M{\"u}ller}, E. 2013{\natexlab{b}}, \prl, 111, 211102,
  \dodoi{10.1103/PhysRevLett.111.211102}

\bibitem[{{Gabler} {et~al.}(2014){Gabler}, {Cerd{\'a}-Dur{\'a}n},
  {Stergioulas}, {Font}, \& {M{\"u}ller}}]{gabler14}
---. 2014, \mnras, 443, 1416, \dodoi{10.1093/mnras/stu1263}

\bibitem[{{Gabler} {et~al.}(2016){Gabler}, {Cerd{\'a}-Dur{\'a}n},
  {Stergioulas}, {Font}, \& {M{\"u}ller}}]{GablerEtAl2016}
---. 2016, \mnras, 460, 4242, \dodoi{10.1093/mnras/stw1272}

\bibitem[{{Hambaryan} {et~al.}(2011){Hambaryan}, {Neuh{\"a}user}, \&
  {Kokkotas}}]{hambaryan}
{Hambaryan}, V., {Neuh{\"a}user}, R., \& {Kokkotas}, K.~D. 2011, \aap, 528,
  A45, \dodoi{10.1051/0004-6361/201015273}

\bibitem[{{Huppenkothen} {et~al.}(2014){Huppenkothen}, {Heil}, {Watts}, \&
  {G{\"o}{\u{g}}{\"u}{\c{s}}}}]{huppen_short}
{Huppenkothen}, D., {Heil}, L.~M., {Watts}, A.~L., \&
  {G{\"o}{\u{g}}{\"u}{\c{s}}}, E. 2014, \apj, 795, 114,
  \dodoi{10.1088/0004-637X/795/2/114}

\bibitem[{{Israel} {et~al.}(2005){Israel}, {Belloni}, {Stella}, {Rephaeli},
  {Gruber}, {Casella}, {Dall'Osso}, {Rea}, {Persic}, \&
  {Rothschild}}]{israel05}
{Israel}, G.~L., {Belloni}, T., {Stella}, L., {et~al.} 2005, \apjl, 628, L53,
  \dodoi{10.1086/432615}

\bibitem[{{Kouveliotou} {et~al.}(1998){Kouveliotou}, {Dieters}, {Strohmayer},
  {van Paradijs}, {Fishman}, {Meegan}, {Hurley}, {Kommers}, {Smith}, {Frail},
  \& {Murakami}}]{kouveliotou98}
{Kouveliotou}, C., {Dieters}, S., {Strohmayer}, T., {et~al.} 1998, \nat, 393,
  235, \dodoi{10.1038/30410}

\bibitem[{{Lander} \& {Jones}(2011)}]{LJ11}
{Lander}, S.~K., \& {Jones}, D.~I. 2011, \mnras, 412, 1730,
  \dodoi{10.1111/j.1365-2966.2010.18009.x}

\bibitem[{{Lander} {et~al.}(2010){Lander}, {Jones}, \& {Passamonti}}]{LJP10}
{Lander}, S.~K., {Jones}, D.~I., \& {Passamonti}, A. 2010, \mnras, 405, 318,
  \dodoi{10.1111/j.1365-2966.2010.16435.x}

\bibitem[{{Lee}(2008)}]{lee08}
{Lee}, U. 2008, \mnras, 385, 2069, \dodoi{10.1111/j.1365-2966.2008.12965.x}

\bibitem[{{Levin}(2007)}]{levin07}
{Levin}, Y. 2007, \mnras, 377, 159, \dodoi{10.1111/j.1365-2966.2007.11582.x}

\bibitem[{{Markey} \& {Tayler}(1973)}]{markeytay73}
{Markey}, P., \& {Tayler}, R.~J. 1973, \mnras, 163, 77,
  \dodoi{10.1093/mnras/163.1.77}

\bibitem[{{Mereghetti} {et~al.}(2006){Mereghetti}, {Esposito}, {Tiengo},
  {Zane}, {Turolla}, {Stella}, {Israel}, {G{\"o}tz}, \& {Feroci}}]{mere06}
{Mereghetti}, S., {Esposito}, P., {Tiengo}, A., {et~al.} 2006, \apj, 653, 1423,
  \dodoi{10.1086/508682}

\bibitem[{{Miller} {et~al.}(2019){Miller}, {Chirenti}, \&
  {Strohmayer}}]{miller19}
{Miller}, M.~C., {Chirenti}, C., \& {Strohmayer}, T.~E. 2019, \apj, 871, 95,
  \dodoi{10.3847/1538-4357/aaf5ce}

\bibitem[{{Nayfeh} \& {Mook}(1979)}]{NayfehMook1979}
{Nayfeh}, A.~H., \& {Mook}, D.~T. 1979, {N}onlinear {O}scillations (New York:
  John Wiley \& Sons).
\newblock \url{http://adsabs.harvard.edu/abs/1979noos.book.....N}

\bibitem[{{Palmer} {et~al.}(2005){Palmer}, {Barthelmy}, {Gehrels}, {Kippen},
  {Cayton}, {Kouveliotou}, {Eichler}, {Wijers}, {Woods}, {Granot}, {Lyubarsky},
  {Ramirez-Ruiz}, {Barbier}, {Chester}, {Cummings}, {Fenimore}, {Finger},
  {Gaensler}, {Hullinger}, {Krimm}, {Markwardt}, {Nousek}, {Parsons}, {Patel},
  {Sakamoto}, {Sato}, {Suzuki}, \& {Tueller}}]{palmer05}
{Palmer}, D.~M., {Barthelmy}, S., {Gehrels}, N., {et~al.} 2005, \nat, 434,
  1107, \dodoi{10.1038/nature03525}

\bibitem[{{Passamonti} \& {Lander}(2013)}]{passland13}
{Passamonti}, A., \& {Lander}, S.~K. 2013, \mnras, 429, 767,
  \dodoi{10.1093/mnras/sts372}

\bibitem[{{Passamonti} \& {Lander}(2014)}]{passland14}
---. 2014, \mnras, 438, 156, \dodoi{10.1093/mnras/stt2134}

\bibitem[{{Pnigouras} \& {Kokkotas}(2015)}]{PnigourasKokkotas2015}
{Pnigouras}, P., \& {Kokkotas}, K.~D. 2015, \prd, 92, 084018,
  \dodoi{10.1103/PhysRevD.92.084018}

\bibitem[{{Pnigouras} \& {Kokkotas}(2016)}]{PnigourasKokkotas2016}
---. 2016, \prd, 94, 024053, \dodoi{10.1103/PhysRevD.94.024053}

\bibitem[{{Pumpe} {et~al.}(2018){Pumpe}, {Gabler}, {Steininger}, \&
  {En{\ss}lin}}]{pumpe18}
{Pumpe}, D., {Gabler}, M., {Steininger}, T., \& {En{\ss}lin}, T.~A. 2018, \aap,
  610, A61, \dodoi{10.1051/0004-6361/201731800}

\bibitem[{{Roberts} {et~al.}(2023){Roberts}, {Baring}, {Huppenkothen},
  {Kouveliotou}, {G{\"o}{\u{g}}{\"u}{\c{s}}}, {Kaneko}, {Lin}, {van der Horst},
  \& {Younes}}]{roberts23}
{Roberts}, O.~J., {Baring}, M.~G., {Huppenkothen}, D., {et~al.} 2023, \apjl,
  956, L27, \dodoi{10.3847/2041-8213/acfcad}

\bibitem[{Sakurai \& Napolitano(2011)}]{SakuraiNapolitano2011}
Sakurai, J.~J., \& Napolitano, J. 2011, {M}odern {Q}uantum {M}echanics, 2nd
  edn. (San Francisco: Addison-Wesley).
\newblock \url{http://adsabs.harvard.edu/abs/1985mqm..book.....S}

\bibitem[{{Samuelsson} \& {Andersson}(2007)}]{sam_and}
{Samuelsson}, L., \& {Andersson}, N. 2007, \mnras, 374, 256,
  \dodoi{10.1111/j.1365-2966.2006.11147.x}

\bibitem[{{Schenk} {et~al.}(2001){Schenk}, {Arras}, {Flanagan}, {Teukolsky}, \&
  {Wasserman}}]{SchenkEtAl2001}
{Schenk}, A.~K., {Arras}, P., {Flanagan}, {\'E}.~{\'E}., {Teukolsky}, S.~A., \&
  {Wasserman}, I. 2001, \prd, 65, 024001, \dodoi{10.1103/PhysRevD.65.024001}

\bibitem[{{Sotani}(2024)}]{sotani_solo_24}
{Sotani}, H. 2024, \prd, 109, 023030, \dodoi{10.1103/PhysRevD.109.023030}

\bibitem[{{Sotani} \& {Kokkotas}(2009)}]{sotani09}
{Sotani}, H., \& {Kokkotas}, K.~D. 2009, \mnras, 395, 1163,
  \dodoi{10.1111/j.1365-2966.2009.14631.x}

\bibitem[{{Sotani} {et~al.}(2007){Sotani}, {Kokkotas}, \&
  {Stergioulas}}]{sotani07}
{Sotani}, H., {Kokkotas}, K.~D., \& {Stergioulas}, N. 2007, \mnras, 375, 261,
  \dodoi{10.1111/j.1365-2966.2006.11304.x}

\bibitem[{{Sotani} {et~al.}(2012){Sotani}, {Nakazato}, {Iida}, \&
  {Oyamatsu}}]{sotani12}
{Sotani}, H., {Nakazato}, K., {Iida}, K., \& {Oyamatsu}, K. 2012, \prl, 108,
  201101, \dodoi{10.1103/PhysRevLett.108.201101}

\bibitem[{{Sotani} {et~al.}(2024){Sotani}, {Suvorov}, \& {Kokkotas}}]{sotani24}
{Sotani}, H., {Suvorov}, A.~G., \& {Kokkotas}, K.~D. 2024, \prd, 110, 023035,
  \dodoi{10.1103/PhysRevD.110.023035}

\bibitem[{{Steiner} \& {Watts}(2009)}]{steiner_watts}
{Steiner}, A.~W., \& {Watts}, A.~L. 2009, \prl, 103, 181101,
  \dodoi{10.1103/PhysRevLett.103.181101}

\bibitem[{{Strohmayer} \& {Watts}(2005)}]{StrohmayerWatts2005}
{Strohmayer}, T.~E., \& {Watts}, A.~L. 2005, \apj, 632, L111,
  \dodoi{10.1086/497911}

\bibitem[{{Strohmayer} \& {Watts}(2006)}]{strohwatts06}
---. 2006, \apj, 653, 593, \dodoi{10.1086/508703}

\bibitem[{{Suvorov} {et~al.}(2022){Suvorov}, {Kuan}, \& {Kokkotas}}]{suv22}
{Suvorov}, A.~G., {Kuan}, H.~J., \& {Kokkotas}, K.~D. 2022, \aap, 664, A177,
  \dodoi{10.1051/0004-6361/202244082}

\bibitem[{{Unno} {et~al.}(1989){Unno}, {Osaki}, {Ando}, {Saio}, \&
  {Shibahashi}}]{UnnoEtAl1989}
{Unno}, W., {Osaki}, Y., {Ando}, H., {Saio}, H., \& {Shibahashi}, H. 1989,
  {N}onradial {O}scillations of {S}tars, 2nd edn. (Tokyo: University of Tokyo
  Press).
\newblock \url{http://adsabs.harvard.edu/abs/1989nos..book.....U}

\bibitem[{{van Hoven} \& {Levin}(2012)}]{vanhoven12}
{van Hoven}, M., \& {Levin}, Y. 2012, \mnras, 420, 3035,
  \dodoi{10.1111/j.1365-2966.2011.20177.x}

\bibitem[{{Watts} \& {Strohmayer}(2006)}]{wattsstroh06}
{Watts}, A.~L., \& {Strohmayer}, T.~E. 2006, \apjl, 637, L117,
  \dodoi{10.1086/500735}

\bibitem[{{Weinberg} {et~al.}(2013){Weinberg}, {Arras}, \&
  {Burkart}}]{WeinbergEtAl2013}
{Weinberg}, N.~N., {Arras}, P., \& {Burkart}, J. 2013, \apj, 769, 121,
  \dodoi{10.1088/0004-637X/769/2/121}

\bibitem[{{Weinberg} {et~al.}(2012){Weinberg}, {Arras}, {Quataert}, \&
  {Burkart}}]{WeinbergEtAl2012}
{Weinberg}, N.~N., {Arras}, P., {Quataert}, E., \& {Burkart}, J. 2012, \apj,
  751, 136, \dodoi{10.1088/0004-637X/751/2/136}

\bibitem[{{Wright}(1973)}]{wright73}
{Wright}, G.~A.~E. 1973, \mnras, 162, 339, \dodoi{10.1093/mnras/162.4.339}

\bibitem[{{Wu} \& {Goldreich}(2001)}]{WuGoldreich2001}
{Wu}, Y., \& {Goldreich}, P. 2001, \apj, 546, 469, \dodoi{10.1086/318234}

\bibitem[{{Xiao} {et~al.}(2024){Xiao}, {Zhang}, {Zhu}, {Xiong}, {Gao}, {Xu},
  {Zhang}, {Peng}, {Li}, {Zhang}, {Lu}, {Lin}, {Liu}, {Zhang}, {Ge}, {Tuo},
  {Xue}, {Fu}, {Liu}, {Liu}, {Li}, {Wang}, {Zheng}, {Wang}, {Jiang}, {Li},
  {Liu}, {Cao}, {Luo}, {Yang}, {Yi}, {Wang}, {Cai}, {Yi}, {Zhao}, {Xie}, {Li},
  {Luo}, {Song}, {Zhang}, {Qu}, {Liu}, {Li}, {Xu}, \& {Li}}]{xiao24}
{Xiao}, S., {Zhang}, Y.-Q., {Zhu}, Z.-P., {et~al.} 2024, \apj, 970, 6,
  \dodoi{10.3847/1538-4357/ad4ee1}

\bibitem[{{Younes} {et~al.}(2017){Younes}, {Baring}, {Kouveliotou}, {Harding},
  {Donovan}, {G{\"o}{\u{g}}{\"u}{\c{s}}}, {Kaspi}, \& {Granot}}]{younes17}
{Younes}, G., {Baring}, M.~G., {Kouveliotou}, C., {et~al.} 2017, \apj, 851, 17,
  \dodoi{10.3847/1538-4357/aa96fd}

\end{thebibliography}
\bibliographystyle{aasjournal}

\end{document}